\documentclass[10pt,conference,anonymous,review]{IEEEtran}

\AtBeginDocument{%
  \providecommand\BibTeX{{%
    \normalfont B\kern-0.5em{\scshape i\kern-0.25em b}\kern-0.8em\TeX}}}

\usepackage[framemethod=tikz]{mdframed} %
\usepackage{enumitem}
\usepackage{stfloats}
\usepackage[numbers]{natbib}
\usepackage[hidelinks]{hyperref}
\usepackage{booktabs}

\usepackage{url}
\usepackage{soul}
\usepackage{colortbl}
\usepackage{adjustbox}
\usepackage{multirow}
\usepackage{tikz}
\usepackage{xcolor}
\usepackage{refcount}
\usepackage{hyperref}
\usepackage{tabularx}
\usepackage{threeparttable}
\usepackage{multicol}
\usepackage{float}
\usepackage{siunitx}
\usepackage{booktabs}
\usepackage{graphicx}

\usepackage[normalem]{ulem}

\definecolor{vlightgray}{gray}{0.9}

\usepackage{hyperref}
\hypersetup{
 colorlinks=true,   
 citecolor=black,
 urlcolor=blue,
 linkcolor=black,
 bookmarksnumbered=true,     
 bookmarksopen=true,         
 bookmarksopenlevel=1,          
 pdfstartview=Fit,           
 pdfpagemode=UseOutlines,    
 pdfpagelayout=TwoPageRight 
}

\newcolumntype{P}[1]{>{\raggedright\arraybackslash}p{#1}}

\newcommand{\boldification}[1]{\ifdraft\indent **\textbf{#1}**\\\indent\else\relax\fi}

\newif\ifdraft
\draftfalse 

\newif\ifrevising
   \revisingfalse 

 \newcommand{\deleted}[1]{{\ifrevising{\relax}\else\relax\fi}}

\begin{document}

\title{What Guides Our Choices? Modeling Developers' Trust and Behavioral Intentions Towards GenAI\vspace{-3mm}}


\author{
\IEEEauthorblockN{
Rudrajit Choudhuri\IEEEauthorrefmark{1}, Bianca Trinkenreich\IEEEauthorrefmark{1}, Rahul Pandita\IEEEauthorrefmark{2}, Eirini Kalliamvakou\IEEEauthorrefmark{2},\\ Igor Steinmacher\IEEEauthorrefmark{3}, Marco Gerosa\IEEEauthorrefmark{3}, Christopher Sanchez\IEEEauthorrefmark{1}, Anita Sarma\IEEEauthorrefmark{1}
	\\
	}
\\
\IEEEauthorblockA{\IEEEauthorrefmark{1}Oregon State University, United States, \{choudhru, bianca.trinkenreich, christopher.sanchez, anita.sarma\}@oregonstate.edu}
\IEEEauthorblockA{\IEEEauthorrefmark{2}GitHub Inc., San Francisco, United States, \{rahulpandita, ikaliam\}@github.com}
\IEEEauthorblockA{\IEEEauthorrefmark{3}Northern Arizona University, United States, \{igor.steinmacher, marco.gerosa\}@nau.edu}
}






%


\newcommand{\explaintwo}[1]{%
\par%
\noindent\fbox{%
    \parbox{\dimexpr\linewidth-2\fboxsep-2\fboxrule}{#1}%
}%
}


\newcommand{\blue}[1]{\textcolor{blue}{#1}}

\maketitle
\IEEEpeerreviewmaketitle

\begin{abstract}

 Generative AI (genAI) tools, such as ChatGPT or Copilot, are advertised to improve developer productivity and are being integrated into software development. However, misaligned trust, skepticism, and usability concerns can impede the adoption of such tools. Research also indicates that AI can be exclusionary, failing to support diverse users adequately. One such aspect of diversity is cognitive diversity---variations in users’ cognitive styles---that leads to divergence in perspectives and interaction styles. When an individual's cognitive style is unsupported, it creates barriers to technology adoption. Therefore, to understand how to effectively integrate genAI tools into software development, it is first important to model 
 \textit{what factors affect developers' trust and intentions to adopt genAI tools in practice?}
%
 
 We developed a theoretically grounded statistical model to (1) identify factors that influence developers’ trust in genAI tools and (2) examine the relationship between developers’ trust, cognitive styles, and their intentions to use these tools in their work. We surveyed software developers (N=238) at two major global tech organizations: GitHub Inc. and Microsoft; and employed Partial Least Squares-Structural Equation Modeling (PLS-SEM) to evaluate our model. Our findings reveal that genAI’s \textit{system/output quality}, \textit{functional value}, and \textit{goal maintenance} significantly influence developers’ trust in these tools. Furthermore, developers' \textit{trust} and \textit{cognitive styles} influence their intentions to use these tools in their work. We offer practical suggestions for designing genAI tools for effective use and inclusive user experience.

\end{abstract}




\section{Introduction}
\label{sec:intro}

Generative AI (genAI) tools (e.g., ChatGPT \cite{GPT4}, Copilot \cite{Copilot}) are being increasingly used in software development \cite{fan2023large}.
These tools promise enhanced productivity \cite{kalliamvakou_2024} and are transforming how developers code and innovate \cite{peng2023impact}. However, this push for adoption \cite{center_2023} is marked with AI hype and skepticism \cite{mckinsey_2024}, as well as interaction challenges \cite{fan2023large, liang2024large}. 

Trust has long been recognized as a critical design requirement of AI tools \cite{hoff2015trust, lee2004trust, sellen2023rise}. Miscalibrated levels of trust---over or under trust---can lead developers to overlook errors and risks introduced by AI \cite{pearce2022asleep} or deter them from using these tools \cite{boubin2017quantifying}. Prior research has identified various factors that foster developers' trust in genAI tools \cite{wang2023investigating, cheng2023would, johnson2023make}. For instance, interaction factors such as setting appropriate expectations and validating AI suggestions \cite{wang2023investigating} along with community factors like shared experiences and community support \cite{cheng2023would} are relevant in building trust. Recently, \citet{johnson2023make} introduced the \textit{PICSE} framework through a qualitative investigation with software developers, outlining key components that influence the formation and evolution of trust in software tools (see Sec. \ref{sec: backg-trust}). What is missing, however, is an empirically grounded theoretical understanding of how the multitude of factors associate with developers' trust in genAI tools.
Therefore, it becomes important to answer \textbf{(RQ1): \textit{What factors predict developers' trust in genAI tools?}} Understanding the significance and strength of these associations is needed to inform the design and adoption of genAI tools in software development.



%
%


Another important concern in industry-wide integration of AI tools is that software design can be exclusionary in different ways \cite{adib2023artificial, beckwith2004gender, business_insider_2023}, often failing to support diverse users \cite{eubanks2018automating}. While a substantial body of work exists on modeling users' technology acceptance \cite{venkatesh2003user, venkatesh2012consumer, chau1996empirical, russo2024navigating}, these studies do not consider the inclusivity of the software design. One often overlooked aspect of inclusivity is supporting cognitive diversity---variations in individuals' cognitive styles---which fosters divergence in perspectives and thoughts (see Sec. \ref{sec: backg-styles}) \cite{sternberg1997cognitive}. Numerous studies have shown that when technology is misaligned with users' diverse cognitive styles \cite{burnett2016gendermag, beckwith2004gender, murphy2024gendermag}, it creates additional barriers for those whose styles are unsupported, forcing them to exert additional cognitive effort \cite{burnett2016gendermag}. Thus, it is essential to understand how developers' cognitive styles influence their intention to adopt genAI tools, and how trust contributes to this multi-faceted decision. Therefore, we investigate \textbf{(RQ2): \textit{How are developers’ trust and cognitive styles associated with their intentions to use genAI tools?}} 

%

We answer these research questions by establishing a theoretical model, grounded in prior literature, for trust and behavioral intentions toward genAI tools. We evaluated this model using Partial Least Squares-Structural Equation Modeling (PLS-SEM) with survey data from developers (N=238) at two global tech organizations: GitHub Inc. \cite{GitHub} and Microsoft\cite{Microsoft}. 


Our theoretical model (Figure~\ref{fig:model}) empirically shows that genAI's \textit{system/output quality} (presentation, adherence to safe and secure practices, performance, and output quality in relation to work style/practices), \textit{functional value} (educational value and practical benefits), and \textit{goal maintenance} (alignment between developers' immediate objectives and genAI's actions) are positively associated with developers' trust in these tools. Furthermore, developers' trust and cognitive styles---intrinsic \textit{motivations} behind using technology, \textit{computer self-efficacy} within peer groups, and \textit{attitudes towards risk}---are associated with their intentions to use these tools, which in turn, correlates with their reported genAI usage in work. 

The main contributions of this paper are twofold: (1) an empirically grounded theoretical model for developers' trust and behavioral intentions towards genAI tools, extending our understanding of AI adoption dynamics in software development, and (2) a psychometrically validated instrument for capturing trust-related factors in the context of human-genAI interactions that can be leveraged in future work. 

\vspace{-2mm}
\vspace{2mm}
\section{Background}
\label{sec: backg}

\vspace{-5px}
\subsection{Trust in AI}
\label{sec: backg-trust}


Trust in AI is commonly defined as \textit{``the attitude that an agent will help achieve an individual's goals in a situation characterized by uncertainty and vulnerability''} \cite{lee2004trust, liao2022designing, vereschak2021evaluate, wang2023investigating, perrig2023trust}. 
%
%
%
Trust is subjective and thus a psychological construct that is not directly observable \cite{hopkins1998educational} and should be distinguished from observable measures such as reliance \cite{wischnewski2023measuring}.
Trust involves users attributing intent and anthropomorphism to the AI \cite{jacovi2021formalizing}, leading to feelings of betrayal when trust is violated. Despite AI systems being inanimate, users often anthropomorphize them \cite{jacovi2021formalizing}, thereby shifting from reliance to trust in AI systems.

Unobservable psychological constructs are commonly measured through validated self-reported scales (instruments) \cite{devellis2021scale} using questions designed to capture the construct of interest.
In this paper, we measure developers' trust in genAI tools using the validated \ul{T}rust in e\ul{X}plainable \ul{AI} (TXAI) instrument \cite{perrig2023trust, hoffman2023measures}. TXAI has been derived from existing trust scales \cite{madsen2000measuring, jian2000foundations, hoffman2023measures} and its psychometric quality has been validated \cite{perrig2023trust}. Researchers frequently advocate using the TXAI instrument for measuring trust in AI \cite{perrig2023trust, scharowski2024trust, makridis2023towards}.


\textit{Factors affecting trust}: Prior research has extensively examined factors influencing human trust in automation \cite{madsen2000measuring, jian2000foundations, mcknight2002developing, merritt2011affective}. However, these preliminary insights do not necessarily transfer to human-AI interactions \cite{wang2023investigating} because of the nuances in how users form trust in AI tools, alongside the inherent uncertainty \cite{vereschak2021evaluate} and variability \cite{weisz2023toward} associated with these systems. Additionally, the context in which AI is applied (in our case, software development) influences how trust is developed and its contributing factors \cite{omrani2022trust}.

Relevant to our domain, \citet{johnson2023make} interviewed software engineers to outline factors that engineers consider when establishing and (re)building trust in tools through the PICSE framework: (1) \textit{\textbf{P}ersonal} (internal, external, and social factors), (2) \textit{\textbf{I}nteraction} (aspects of engagement with a tool), (3) \textit{\textbf{C}ontrol} (over the tool), (4) \textit{\textbf{S}ystem} (properties of the tool), and (5) \textit{\textbf{E}xpectations} (with the tool). Since PICSE is developed for software engineering (SE), we use it to design our survey instrument to identify factors influencing developers' trust in genAI tools. However, the PICSE framework was qualitatively developed and the psychometric quality---reliability and validity---of a survey based on it has not yet been assessed. 

Our work builds upon PICSE to contribute (a) a validated instrument for capturing different factors that developers consider when forming trust in genAI tools (Sec. \ref{sec:psycho}) through a psychometric analysis of the PICSE framework and (b) assesses the significance and strength of these factors' association with trust in genAI tools (Sec. \ref{sec:Res}).



\vspace{-2mm}
\subsection{Users' Cognitive Styles}
\label{sec: backg-styles}
AI can be exclusionary in different ways
often failing to support all users as it should \cite{eubanks2018automating, adib2023artificial, business_insider_2023}. E.g., \citet{weisz2022better} found that some, but not all, participants could produce high-quality code with AI assistance, and the differences were linked to varying participant interactions with AI.

User experience in Human-AI interaction (HAI-UX) can be improved by supporting diverse cognitive styles \cite{anderson2022measuring}, which refer to the \textit{ways users perceive, process, and interact with information and technology, as well as their approach to problem-solving} \cite{sternberg1997cognitive}. While no particular style is inherently better or worse, if a tool insufficiently supports (or is misaligned with) users' cognitive styles; they pay an additional ``cognitive tax'' to use it, creating barriers to usability \cite{murphy2024gendermag}.

Here, we scope developers' diverse cognitive styles to the five cognitive styles in the GenderMag inclusive design method \cite{burnett2016gendermag}. GenderMag's cognitive styles (facets) are users' diverse: \textit{attitudes towards risk}, \textit{computer self-efficacy} within their peer group, \textit{motivations} to use the technology, \textit{information processing style}, and \textit{learning style} for new technology.
Each facet represents a spectrum. For example, risk-averse individuals (one endpoint of the `attitude towards risk' spectrum) hesitate to try new technology or features, whereas risk-tolerant ones (the other end) are inclined to try unproven technology that may require additional cognitive effort or time. GenderMag's cognitive styles are well-suited as they have been (a) repeatedly shown to align with users' interactions with technology both in the context of SE \cite{murphy2024gendermag, burnett2016gendermag, guizani2022debug} and HAI interactions \cite{anderson2022measuring, hamid2024improving}, and (b) distilled from an extensive list of applicable cognitive style types \cite{beckwith2004gender, burnett2016gendermag}, intended for actionable use by practitioners. 
We used the validated GenderMag facet survey instrument \cite{hamid2023measure} in our study. 

\vspace{-1mm}
\subsection{Behavioral Intention and Usage}
\label{sec: backg-behavioral}

Behavioral intention refers to \textit{the extent to which a person has made conscious plans to undertake a specific future activity} \cite{venkatesh2003user}. Technology acceptance models, such as TAM \cite{chau1996empirical} and UTAUT \cite{venkatesh2003user}, identify behavioral intention as a key indicator of actual technology usage \cite{venkatesh2012consumer}. Understanding users’ behavioral intentions is useful for predicting technology adoption and guiding future design strategies \cite{venkatesh2003user}. While there is an extensive body of work modeling users' behavioral intentions towards software tools \cite{venkatesh2003user, venkatesh2012consumer, russo2024navigating}, these studies primarily focus on socio-technical factors driving adoption.

Our work contributes to this line of research by examining the role of developers’ trust and cognitive styles in shaping their intentions to use genAI tools (Sec. \ref{sec:theory} and \ref{sec:Res}), thereby extending the understanding of AI adoption dynamics in SE. We used components of the UTAUT model \cite{venkatesh2003user} to capture developers’ behavioral intentions and usage of genAI tools.

\vspace{-2mm}
\vspace{1mm}
\section{Method} 
\label{sec:method}

To address our RQs, we surveyed software developers from two major global tech organizations, GitHub Inc. and Microsoft. We leveraged existing theoretical frameworks and instruments to design our data collection instrument (see Sec. \ref{sec: backg}). While using existing theoretical frameworks is a first step in developing questionnaires, conducting a psychometric quality assessment is essential to ensure its subsequent reliability and validity \cite{furr2011scale}. As there was no validated instrument to measure the constructs of the PICSE framework \cite{johnson2023make}--our chosen trust framework--we performed its psychometric assessment \cite{furr2011scale} (Sec. \ref{sec:psycho}). This assessment helped us define a theoretical model of factors developers consider when forming trust in genAI tools, which we then evaluated using Partial Least Squares-Structural Equation Modeling (PLS-SEM) to answer RQ1. To answer RQ2, we assessed the relationships between developers' trust and cognitive styles with their intentions to use genAI tools. 
Next, we discuss each step.

\subsection{Survey Design and Data Collection}
\label{sec:datacollection}

\subsubsection{\textbf{Survey design}} 
We defined the measurement model \cite{hair2019use} based on the theoretical frameworks discussed in Sec. \ref{sec: backg}  to guide our survey design (Table \ref{tab:mm-instruments}). 
Four researchers with experience in survey studies and GitHub's research team co-designed the survey over a four-month period (Oct 2023 to Jan 2024). We adapted existing (validated) instruments in designing the survey questions (Table \ref{tab:mm-instruments}).
The questions were contextualized for the target population and pragmatic decisions were made to limit the survey length.
The complete questionnaire is available in the supplemental material \cite{supplemental}.


\vspace{-3mm}
\begin{table}[!ht]
\caption{Measurement Model Constructs and Instruments}
\label{tab:mm-instruments}
\vspace{-8px}
\centering
\begin{tabular}{>{\raggedright\arraybackslash}m{4cm} >{\raggedright\arraybackslash}m{4cm}}
\hline
\textbf{Construct} & \textbf{Instrument} \\
\hline \hline
\rowcolor{gray!25} Trust & TXAI instrument* \cite{perrig2023trust, hoffman2023measures} \\ 
Factors affecting trust & PICSE framework** \cite{johnson2023make} \\
\rowcolor{gray!25} Users' cognitive styles & GenderMag facet survey \cite{hamid2023measure} \\
Behavioral intention \& usage & UTAUT model \cite{venkatesh2003user, venkatesh2012consumer} \\
\hline
\end{tabular}

\begin{tablenotes}
  \scriptsize
  \item
*We used the 4-item TXAI scale \cite{hoffman2023measures} instead of the 6-item scale \cite{perrig2023trust} to reduce participant fatigue. **PICSE does not have a validated questionnaire in \cite{johnson2023make}.
\vspace{-5px}
\end{tablenotes}
\end{table}

After the IRB-approved informed consent, participants responded to closed questions about their familiarity with genAI technology and their attitudes and intentions towards using genAI tools in work.  All closed questions utilized a 5-point Likert scale ranging from 1 (``strongly disagree'') to 5 (``strongly agree'') with a neutral option. These questions also included a $6^{th}$ option (``I'm not sure'') for participants who either preferred not to or did not know how to respond to a question. This differs from being neutral--acknowledging the difference between ignorance and indifference \cite{grichting1994meaning}. 

Demographic questions covered gender, continent of residence, years of software engineering (SE) experience, and primary SE responsibilities at work. We did not collect data on country of residence or specific job roles/work contexts to maintain participant anonymity, as per GitHub and Microsoft's guidelines. An open-ended question for additional comments was included at the end of the survey. 

The survey took between 7-10 minutes to complete. Attention checks were included to ensure the quality of the survey data.
To reduce response bias, we randomized the order of questions within their respective blocks (each construct in Table \ref{tab:mm-instruments}). 
We piloted the questionnaire with collaborators at GitHub to refine its clarity and phrasing.



\begin{table}[!t]
\scriptsize
\centering
\caption{Respondent Demographics (n=238)}
\label{tab:demographics}
\vspace{-8px}
\robustify{\bfseries}
\begin{tabular}{p{5.8cm}
                S[table-format=3.0]
                S[table-format=2.1]}
\toprule
Attribute & {N} & {Percentage}\\
\midrule
\multicolumn{3}{c}{\textbf{Gender}}\\
\midrule
\rowcolor{gray!25} Man & 186 & 78.2\si{\percent}\\
Woman & 39 & 16.4\%\\
\rowcolor{gray!25} Non-binary or gender diverse & 6 & 2.5\%\\
Prefer not to say & 7 & 2.9\%\\

\midrule
\multicolumn{3}{c}{\textbf{Continent of Residence}}\\
\midrule
\rowcolor{gray!25} North America & 129 & 54.2\%\\
Europe & 55 & 23.1\%\\
\rowcolor{gray!25} Asia & 33 & 13.8\%\\
Africa & 9 & 3.8\%\\
\rowcolor{gray!25} South America & 8 &3.4\%\\
Pacific/Oceania & 4 & 1.7\%\\

\midrule
\multicolumn{3}{c}{\textbf{SE Experience}}\\
\midrule
\rowcolor{gray!25} 1-5 years & 57 & 23.9\%\\
6-10 years & 50 & 21.0\%\\
\rowcolor{gray!25} 11-15 years & 52 & 21.9\%\\
Over 15 years & 79 & 33.2\%\\

\midrule
\multicolumn{3}{c}{\textbf{SE Responsibilities}}\\
\midrule
\rowcolor{gray!25} Coding/Programming & 223 & 93.7\%\\
Code Review & 192 & 80.6\%\\
\rowcolor{gray!25} System Design & 148 & 62.1\%\\
Documentation & 110 & 46.2\%\\
\rowcolor{gray!25} Maintenance \& Updates & 108 & 45.4\%\\
Requirements Gathering \& Analysis & 108 & 45.4\%\\
\rowcolor{gray!25} Performance Optimization & 107 & 44.9\%\\
Testing \& Quality Assurance & 98 & 41.2\%\\
\rowcolor{gray!25} DevOps/(CI/CD) & 90 & 37.8\%\\
Project Management \& Planning & 53 & 22.3\%\\
\rowcolor{gray!25} Security Review \& Implementation & 46 & 19.3\%\\
Client/Stakeholder Communication & 32 & 13.5\%\\

\bottomrule
\end{tabular}
\vspace{-6mm}
\end{table}

\subsubsection{\textbf{Distribution}} 
GitHub and Microsoft administered the online questionnaire using their internal survey tools. The survey was distributed to team leads, who were asked to cascade it to their team members. This approach was chosen over using mailing lists to ensure a broader reach \cite{trinkenreich2023belong}. The survey was available for one month (Feb-Mar, 2024), and while participation was optional, it was encouraged.  

\subsubsection{\textbf{Responses}} 
We received a total of 343 responses: 235 from Microsoft and 108 from GitHub. We removed patterned responses (n=20), outliers ($<1$ year SE experience, n=1), and those that failed attention checks (n=29). Further, we excluded respondents who discontinued the survey without answering all the close-ended questions (n=55). We considered ``I'm not sure'' responses as missing data points. As in prior work \cite{trinkenreich2023belong}, we did not impute data points due to the unproven efficacy of imputation methods within SEM group contexts \cite{sarstedt2017treating}.

After filtration, we retained 238 valid responses (Microsoft: 154, GitHub: 84) from developers across six continents, representing a wide distribution of SE experience. Most respondents were from North America (54.2\%) and Europe (23.1\%), and most identified as men (78.2\%), aligning with distributions reported in previous studies with software engineers \cite{trinkenreich2023belong, russo2024navigating}. Table \ref{tab:demographics} summarizes the respondent demographics. 


\subsection{Psychometric Analysis of PICSE Framework}
\label{sec:psycho}

Psychometric quality \cite{lord2008statistical, raykov2011introduction} refers to the objectivity, reliability, and validity of an instrument. We primarily used validated instruments in designing the survey. However, since PICSE was not validated, we conducted a psychometric analysis to empirically refine its factor groupings, which were then evaluated for their association with trust (Sec. \ref{sec:Res}). Table \ref{tab:categories_items} presents the factors evaluated in our survey. We performed the analysis using the JASP tool \cite{jasp_website}, adhering to established psychometric procedures \cite{raykov2011introduction, howard2016review, perrig2023trust} as detailed below:

%
%


\textit{1) \textbf{Confirmatory Factor Analysis (CFA) -- Original grouping}}: CFA is a statistical technique that examines intercorrelations between items and proposed factors to test whether a set of observed variables align with a pre-determined factor structure~\cite{harrington2009confirmatory}. We assessed whether the PICSE items align with their original five-factor structure (Personal, Interaction, Control, System, and Expectations). The model fit was evaluated using multiple indices: Chi-square test, Root Mean Square Error of Approximation (RMSEA), Comparative Fit Index (CFI), and Tucker-Lewis Index (TLI) \cite{hu1999cutoff}. Indications of a good model fit include 
$p>.05$ for $\chi^2$ test, $\text{RMSEA}<.06$, $\text{SRMR} \leq .08$, and $0.95 \leq \text{CFI, TLI} \leq 1$ \cite{hu1999cutoff}.
We employed robust maximum likelihood estimator (MLE),\footnote{Robust maximum likelihood estimation adjusts for non-normality in data. In general, ``robust" in factor analyses refer to methods and values resilient to deviations from ideal distributional assumptions.} since the data did not meet multivariate normality assumptions, confirmed using Mardia’s test \cite{mardia1970measures} (see supplemental \cite{supplemental}).
As shown in Table \ref{tab:fit_indices}, results from the original five-factor structure did not indicate a good model fit based on RMSEA, SRMR, CFI, and TLI. This was not entirely unexpected given PICSE's conceptual nature \cite{harrington2009confirmatory}.
Therefore, to identify a more appropriate model of factors, we proceeded with an exploratory factor analysis (EFA), uncovering alternative groupings that might better fit the data.

\vspace{-3mm}
\begin{table}[bhtp]
\caption{PICSE framework \cite{johnson2023make}}
\label{tab:categories_items}
\vspace{-10pt}
\centering
\begin{tabular}{>{\raggedright\arraybackslash}m{1.5cm} >{\raggedright\arraybackslash}m{6.5cm}}
\hline
\textit{\textbf{Category}} & \textit{\textbf{Items}} \\
\hline \hline
\rowcolor{gray!25} Personal & Community (P1), Source reputation (P2), Clear advantages (P3) \\ 
Interaction & Output validation support (I1), Feedback loop (I2), Educational value (I3) \\
\rowcolor{gray!25} Control & Control over output use (C1), Ease of workflow integration (C2) \\
System & Ease of use (S1), Polished presentation (S2), Safe and secure practices (S3), Consistent accuracy and appropriateness (S4), Performance (S5) \\
\rowcolor{gray!25} Expectations & Meeting expectations (E1), Transparent data practices (E2), Style matching (E3), Goal maintenance (E4) \\
\hline
\end{tabular}
\begin{tablenotes}
\scriptsize
\item We dropped C3 (tool ownership), as it pertained to AI engineers developing parts of genAI models.
\vspace{-15px}
\end{tablenotes}
\end{table}


\vspace{3mm}

\textit{2) \textbf{Exploratory Factor Analysis (EFA)}}: Unlike CFA, which relies on an existing a priori expectation of factor structures, EFA identifies the suitable number of latent constructs (factors) and underlying factor structures without imposing a preconceived model \cite{howard2016review}. Given the violation of multivariate normality, we used principal axis factoring (considering factors with eigenvalues$>1$), as recommended for EFA \cite{howard2016review}. We employed oblique rotation, anticipating correlations among the factors. The data met EFA assumptions: significant Bartlett’s test for sphericity ($\chi^2$ (136) = 1633.97, $p < .001$) and an adequate Kaiser-Meyer-Olkin (KMO) test (0.892, $\text{recommended} \geq .60$) \cite{howard2016review}. 
We used both parallel analysis and a scree plot to determine the number of factors \cite{howard2016review}, suggesting an alternate five-factor model explaining 64.6\% of the total variance. However, most of this variance (60.3\%) was accounted for by factors 1, 2, 3, and 5 (22.7, 15.1, 10.6, and 11.9\% respectively), while factor 4 explained only 4.3\%. The factors showed low correlations (0.2-0.3), except for factor 4, which had high correlations with factors 2 (0.625) and 3 (0.524) (see supplemental). Table \ref{tab:factor_loadings} presents the factor loadings, which indicate the extent to which changes in the underlying factor are reflected in the corresponding indicator (item). All items loaded well onto their factors with primary $\text{loadings} > 0.5$ (items with loadings below 0.4 should be excluded) \cite{hair2009multivariate, hair2019use}. 
For interpreting communality (i.e., the proportion of an item's variance explained by the common factors in factor analyses), $\text{values} < 0.5$ are considered problematic and not interpretable~\cite{hair2009multivariate}. As shown in Table~\ref{tab:factor_loadings}, the communality values for items I1, I2, E1, and E2 were below 0.5, and as they did not load onto any of the factors, they were excluded from the final model. Items in factor 4 (P1, P2, C1) also had low communality values and were likewise dismissed. Based on these results, we concluded that a four-factor solution was the most appropriate, dropping factor 4 due to its low variance explanation, high correlations with other factors, and low communality. The fit indices in Table \ref{tab:fit_indices} indicate a good model fit, showing that the EFA factor structure better fits the data than the original PICSE grouping in the above CFA analysis.

\vspace{-6mm}
\begin{table}[htbp]
\scriptsize
\caption{EFA: Factor loadings and communalities ($h2$)}
\label{tab:factor_loadings}
\vspace{-8pt}
\centering
\begin{tabular}{>{\raggedright\arraybackslash}m{0.3cm} >{\centering\arraybackslash}m{1cm} >{\centering\arraybackslash}m{1cm} >{\centering\arraybackslash}m{1cm} >{\centering\arraybackslash}m{1cm} >{\centering\arraybackslash}m{1cm} >{\centering\arraybackslash}m{0.7cm}}
\hline
Item & Factor 1 & Factor 2 & Factor 3 & Factor 4 & Factor 5 & $h2$ \\
\hline \hline
\rowcolor{gray!30} S2 & 0.655 & & & & & 0.525 \\
S3 & 0.729 & & & & & 0.609 \\
\rowcolor{gray!30} S4 & 0.823 & & & & & 0.623 \\
S5 & 0.638 & & & & & 0.657 \\
\rowcolor{gray!30} E3 & 0.614 & & & & & 0.559 \\
I3 & & 0.941 & & & & 0.779 \\
\rowcolor{gray!30} P3 & & 0.791 & & & & 0.599 \\
S1 & & & 0.739 & & & 0.668 \\
\rowcolor{gray!30} C2 & & & 0.671 & & & 0.607 \\
P1 & & & & 0.613 & & \textcolor{red}{0.418} \\
\rowcolor{gray!30} P2 & & & & 0.539 & & \textcolor{red}{0.367} \\
C1 & & & & 0.517 & & \textcolor{red}{0.312} \\
\rowcolor{gray!30} E4 & & & & & 0.628 & 0.685 \\
I1 & & & & & & \textcolor{red}{0.398} \\
\rowcolor{gray!30} I2 & & & & & & \textcolor{red}{0.481} \\
E1 & & & & & & \textcolor{red}{0.405} \\
\rowcolor{gray!30} E2 & & & & & & \textcolor{red}{0.492} \\
\hline
\end{tabular}
\begin{tablenotes}
\item The applied oblique rotation method is promax. Communality values (h2)\textless 0.5 are problematic \cite{hair2009multivariate} and are marked in \textcolor{red}{RED}. Items loaded well (\textgreater 0.5) onto their primary factors without cross-loadings (\textgreater 0.3) onto other factors \cite{hair2019use}; hence their corresponding cells are kept blank.
\vspace{-10pt}
    \end{tablenotes}
\end{table}

\textit{3) \textbf{CFA - Alternate grouping}}: In the final step, as is best practice \cite{raykov2011introduction}, we conducted CFA to validate the factor structure identified through EFA. The CFA fit indices in Table~\ref{tab:fit_indices} confirm the EFA-derived four-factor model (RMSEA = 0.048; SRMR = 0.047, CFI = 0.982, TLI = 0.973).
Table~\ref{tab:factor_loadings} outlines the factor structure and corresponding item groupings. Factor 1, labeled \textbf{\textit{System/Output quality}}, includes items S2 through S5 and E3, which relate to the System group (in PICSE) and the style matching of genAI's outputs. Factor 2, labeled \textit{\textbf{Functional value}}, encompasses items I3 and P3, reflecting the educational value and practical advantages of using genAI tools. Factor 3, labeled \textbf{\textit{Ease of use}}, comprises items S1 and C2, addressing the ease of using and integrating genAI in the workflow. Factor 5, labeled \textbf{\textit{Goal maintenance}}, includes a single item, E4, focusing on genAI's maintenance of human goals. The reliability and validity assessments further support the robustness of these constructs (see Sec. \ref{mm-eval}). 

In summary, the psychometric analysis confirmed that a four-factor solution is most appropriate and provided a validated measurement instrument for capturing these factors.

\vspace{-3mm}
\begin{table}[!ht]
\footnotesize
\caption{Model Fit Indices - PICSE Psychometric Evaluation}
\label{tab:fit_indices}
\vspace{-8pt}
\centering
\begin{tabular}{>{\raggedright\arraybackslash}m{1.7cm} >{\centering\arraybackslash}m{0.8cm} >{\centering\arraybackslash}m{0.8cm} >{\centering\arraybackslash}m{0.6cm} >{\centering\arraybackslash}m{0.6cm} >{\centering\arraybackslash}m{0.6cm} >{\centering\arraybackslash}m{0.7cm}}
\hline
Model & \textit{RMSEA} & \textit{SRMR }& \textit{CFI} & \textit{TLI} & \textit{$\chi^2$} & \textit{p-val} \\
\hline \hline
\rowcolor{gray!25} CFA-Original & 0.104 & 0.084 & 0.925 & 0.927 & 147.3 & $<$0.01 \\
EFA & 0.057 & 0.054 & 0.968 & 0.965 & 109.1 & $<$0.01 \\ 
\rowcolor{gray!25} CFA-Alternate & 0.048 & 0.047 & 0.982 & 0.973 & 59.0 & $<$0.01 \\
\hline
\end{tabular}
\begin{tablenotes}
  \scriptsize
  \item
$\chi^2$ test results were not considered, as the test is affected by deviations from multivariate normality \cite{schumacker2004beginner}. We still report the values for completeness.
\vspace{-17px}
\end{tablenotes}
\end{table}



  



\vspace{2mm}
\subsection{Model Development}
\label{sec:theory}

As discussed in the previous section, we refined the factor groupings within the PICSE framework. In this study, we are not proposing fundamentally different relationships to trust beyond those identified in the PICSE framework. Instead, we have constrained our focus to only those factors that were psychometrically validated. Next, we detail the hypotheses embedded in our theoretical model for each research question.



\vspace{0.5mm} 
\noindent \textit{\textbf{RQ1)} Factors associated with trust}
\vspace{0.5mm} 


\textbf{\textit{System/Output quality}} encompasses genAI tools' presentation, adherence to safe and secure practices (including privacy and security implications of using genAI), and its performance and output quality (consistency and correctness) in relation to the development style or work environment in which it is utilized (S2-S5, E3). 
Developers often place trust in AI based on its performance and output quality (accuracy and consistency), which serve as proxies for the system's perceived credibility \cite{fogg1999elements, wang2023investigating, cheng2023would, yu2019trust}. 
Prior work \cite{wang2023investigating} evidenced that developers are often wary about the security and privacy implications of using AI tools in their work, which influences the level of trust they place in these tools. Drawing upon these insights, we hypothesize: 
\textit{\textbf{(H1)} System/Output quality of genAI is positively associated with developers' trust in these tools.}

\textbf{\textit{Functional value}} of a tool refers to the practical benefits and utility it offers users in their work \cite{sheth1991we}. In our context, genAI's functional value encompasses its educational value and clear advantages relative to work performance (I3, P3). Prior work highlights that developers' expectations of clear advantages from using AI tools (e.g., increased productivity, improved code quality) contribute to their trust in using these tools \cite{johnson2023make, ziegler2024measuring}. Further, AI's ability to support learning fosters trust in these tools \cite{wang2023investigating}. Based on these, we posit: 
\textit{\textbf{(H2)} Functional value of genAI is positively associated with developers' trust in these tools.}

\textbf{\textit{Ease of use}} associated with genAI tools includes the extent to which developers can easily use and integrate genAI into their current workflow (S1, C2). Prior research highlights that a tool's ease of use \cite{gefen2003trust} and compatibility with existing workflows~\cite{lee2004trust, russo2024navigating} contribute to users' trust. Following this, we hypothesize:
\textit{\textbf{(H3)} GenAI's ease of use is positively associated with developers' trust in these tools.}

\textbf{\textit{Goal maintenance}} is related to the degree to which genAI's actions and responses align with the developer's ongoing goals (E4). By its very nature, goals can vary depending on the task and context \cite{johnson2023make}. Therefore, aligning AI behavior with an individual’s immediate goals is crucial in human-AI collaboration scenarios \cite{wischnewski2023measuring}. In terms of human cognition, this congruence is important for maintaining cognitive flow and reducing cognitive load \cite{unsworth2012variation}, which, in turn, fosters trust in systems \cite{van2019trust, chow2008social}. 
Consequently, we propose:
\textit{\textbf{(H4)} Goal maintenance is positively associated with developers' trust in genAI tools.}

\vspace{0.5mm}
\noindent \textit{\textbf{RQ2)} Factors associated with behavioral intentions}
\vspace{0.5mm}

\textbf{\textit{Trust}} is a key factor in explaining resistance toward automated systems \cite{wischnewski2023measuring} and plays an important role in technology adoption \cite{xiao2014social, witschey2015quantifying}.
Multiple studies have correlated an individual's trust in technology with their intention to use it \cite{kim2019study, gefen2003trust, baek2023chatgpt}. 
In our context, we thus posit:
\textit{\textbf{(H5)} Trust is positively associated with intentions to use genAI tools.}


\noindent In the context of GenderMag's cognitive styles:

\textbf{\textit{Motivations}} behind why someone uses technology (technophilic or task-focused) not only influences their intention to use it but also affects how they engage with its features and functionalities \cite{venkatesh2012consumer, o2010influence}.
Naturally, individuals motivated by their interest and enjoyment in using and exploring the technology (opposite end of the spectrum from those motivated by task completion) are early adopters of new technology \cite{burnett2016gendermag}.
 Based on this, we posit:
%
\textit{\textbf{(H6)} Motivation to use technology for its own sake is positively associated with intentions to use genAI tools.}

\textbf{Computer self-efficacy} refers to an individual's belief in their ability to engage with and use new technologies to succeed in tasks \cite{bandura1997self}. It shapes how individuals apply cognitive strategies and the effort and persistence they invest in using new technologies \cite{compeau1995computer}, 
thereby influencing their intention to use them \cite{venkatesh2000theoretical, li2024effect}. In line with this, we propose:
\textit{\textbf{(H7)} Computer self-efficacy is positively associated with intentions to use genAI tools.}

\textbf{Attitude towards risk} encompasses an individual's inclination to take risks in uncertain outcomes \cite{byrnes1999gender}. This cognitive facet influences decision-making processes, particularly in contexts involving new or unfamiliar technology \cite{venkatesh2000theoretical}. Risk-tolerant individuals (one end of the spectrum) are more inclined to experiment with unproven technology than risk-averse ones (the other end) \cite{burnett2016gendermag}, and show higher intentions to use new tools \cite{venkatesh2012consumer, li2024effect}. Thus, we posit:
\textit{\textbf{(H8)} Risk tolerance is positively associated with intentions to use genAI tools.}

\textbf{Information processing} style influences how individuals interact with technology when problem-solving: 
some gather information \textit{comprehensively} 
to develop a detailed plan before acting; others gather information \textit{selectively}, acting on initial promising pieces and acquiring more as needed \cite{burnett2016gendermag}. GenAI systems, by their very interaction paradigm, inherently support the latter by providing
immediate responses to queries, allowing users to act quickly on the information received and gather additional details incrementally. Accordingly, we posit:
\textit{\textbf{(H9)} Selective information processing style is positively associated with intentions to use genAI tools.}

\textbf{Learning style for technology} (by process vs. by tinkering) refers to how an individual approaches problem-solving and how they structure their approach to a new technology \cite{burnett2016gendermag}. Some prefer to learn through an organized, step-by-step process, while others prefer to tinker around—exploring and experimenting with new technology or its features \cite{burnett2016gendermag}. Prior work indicates that software, more often than not, is designed to support and encourage tinkering \cite{carroll2003design}, making individuals who prefer this approach more inclined to adopt and use new tools \cite{venkatesh2003user}. 
Thus, we propose:
\textit{\textbf{(H10)} Tinkering style is positively associated with intentions to use genAI tools.}

\boldification{Intention to actual use}

\textbf{\textit{Behavioral intention.}} Successful technology adoption hinges on users' intention to use it, translating into future usage. Prior work has consistently shown these factors to be positively correlated \cite{venkatesh2003user, venkatesh2012consumer}, suggesting that users who intend to use technology are more likely to do so. Accordingly, we hypothesize: 
\textit{\textbf{(H11)} Behavioral intention to use genAI tools is positively associated with the usage of these tools.}


%

\vspace{-1mm}
\subsection{Data Analysis}
\label{instr-design}

We used Partial Least Squares-Structural Equation Modeling (PLS-SEM) to test our theoretical model. PLS-SEM is a second-generation multivariate data analysis technique that has gained traction in empirical SE studies investigating complex phenomena \cite{russo2024navigating, trinkenreich2023belong, russo2021pls}.
It allows for simultaneous analysis of relationships among constructs (measured by one or more indicators) and addresses multiple interconnected research queries in one comprehensive analysis. 
It is particularly suited for exploratory studies due to its flexibility in handling model complexity while accounting for measurement errors in latent variables \cite{hair2019use}. Importantly, PLS-SEM does not require data to meet distributional assumptions. Instead, it uses a bootstrapping approach to determine the statistical significance of path coefficients (i.e., relationships between constructs). The PLS path model is estimated for a large number of random subsamples (usually 5000), generating a bootstrap distribution, which is then used to make statistical inferences \cite{hair2019use}.


We used the SmartPLS (v4.1.0) software \cite{smartpls_website} for PLS-SEM analyses, which comprised two main steps, each involving specific tests and procedures. First, we evaluated the measurement model, empirically assessing the relationships between the latent constructs and their indicators (Sec. \ref{mm-eval}). Next, we evaluated the theoretical (or structural) model (Sec. \ref{str-eval}), representing the hypotheses presented in Section \ref{sec:theory}. 

The appropriate sample size was determined by conducting power analysis using the G*Power tool ~\cite{faul2009statistical}. We performed an \textit{F}-test with multiple linear regression, setting a medium effect size (0.25), a significance level of 0.05, and a power of 0.95. The maximum number of predictors in our model is seven (six theoretical constructs and one control variable to Behavioral Intention) (see Fig.~\ref{fig:model}). The calculation indicated a minimum sample size of 95; our final sample size of 238 exceeded it considerably.

\vspace{-2mm}
\vspace{2mm}
\section{Results}
\label{sec:Res}

In this section, we report the evaluation of the measurement model (Sec.~\ref{mm-eval}), followed by the evaluation of the structural model (Sec.~\ref{str-eval}). We adhered to the evaluation protocols outlined in prior studies \cite{russo2021pls, hair2019use}. 
%
%
The analysis was performed using the survey data, which met the assumptions for factor analysis \cite{hair2019use}: significant Bartlett’s test of sphericity on all constructs ($\chi^2$(496)=4474.58, p $<$ .001) and adequate KMO measure of sampling adequacy (0.901), well above the recommended threshold (0.60) \cite{howard2016review}. 

\vspace{-2mm}
\subsection{Measurement Model Evaluation}
\label{mm-eval}

\begin{figure*}[!bht]
\centering
\includegraphics[width=0.9\textwidth]{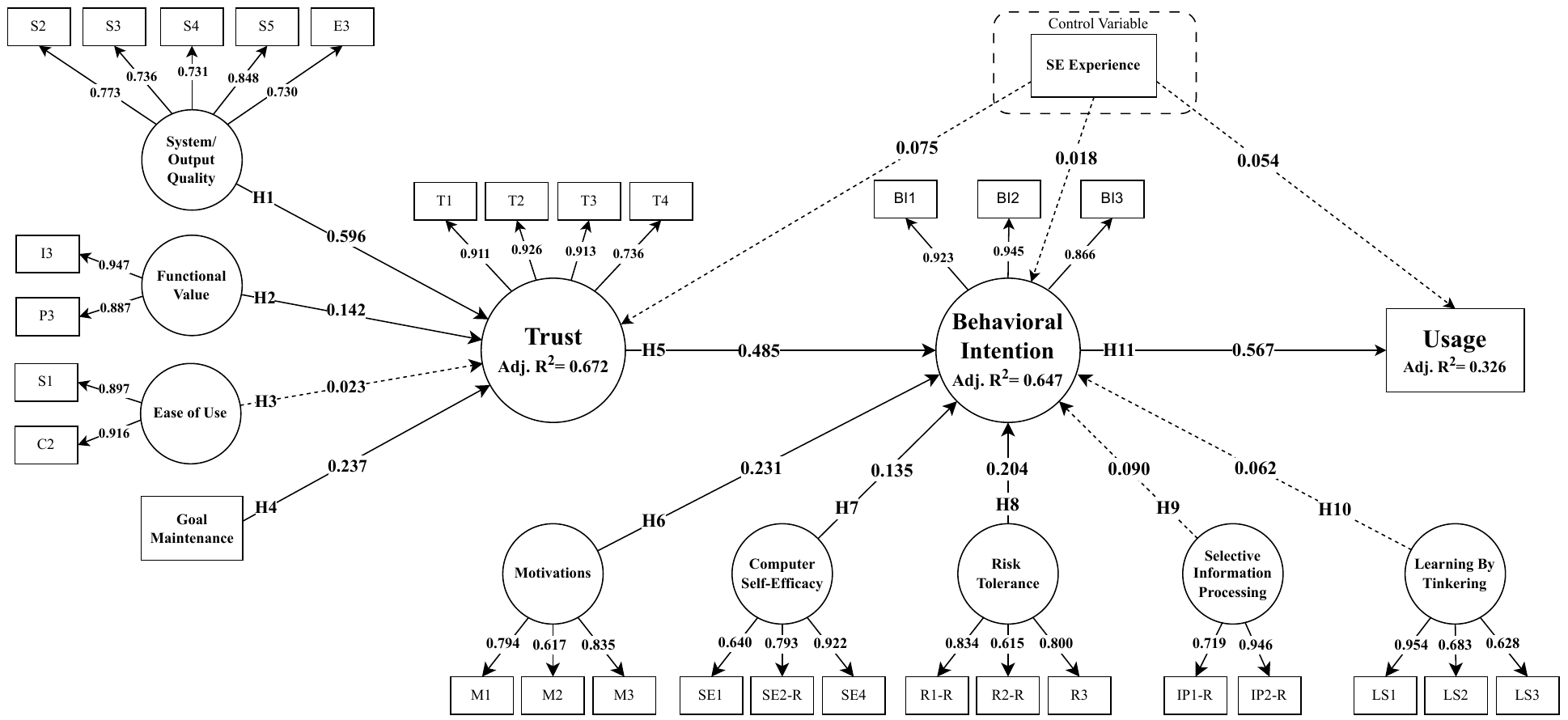}
\vspace{-5px}
\caption{PLS-SEM Model: Solid lines indicate item loadings and path coefficients (p $<$ 0.05); dashed lines represent non-significant paths. Reverse-coded items are suffixed with `-R' (e.g., SE2-R). Latent constructs are depicted as circles and adjusted $R^2$ (Adj. $R^2$) values are reported for endogenous constructs.}
\vspace{-5px}
\label{fig:model}
\vspace{-10px}
\end{figure*}

Our model evaluates several theoretical constructs that are not directly observable (e.g., Trust, Behavioral Intention). These constructs are modeled as latent variables, each measured by a set of indicators or manifest variables (see Fig. \ref{fig:model}). The first step in evaluating a structural equation model is to ensure the soundness of the measurement of these latent variables, a process referred to as evaluating the `measurement model'~\cite{hair2019use}. We performed a series of tests to validate the measurement model \cite{russo2021pls}, detailed as follows:

\vspace{-3mm}
\begin{table}[h]
\centering
\caption{Internal Consistency Reliability and Convergent Validity}
\vspace{-8px}
\begin{tabular}{>{\raggedright\arraybackslash}m{2.65cm} >{\centering\arraybackslash}m{1.8cm} >{\centering\arraybackslash}m{0.8cm} >{\centering\arraybackslash}m{0.8cm} >{\centering\arraybackslash}m{0.8cm}}

\hline
\textbf{} & \textit{Cronbach's $\alpha$} & \textit{CR($\rho_a$)} & \textit{CR($\rho_c$)} & \textit{AVE} \\ \hline \hline
\rowcolor{gray!25} System/Output quality & 0.816 & 0.834 & 0.874 & 0.781 \\
Functional value & 0.816 & 0.895 & 0.914 & 0.842 \\
\rowcolor{gray!25} Ease of use & 0.780 & 0.782 & 0.902 & 0.822 \\
Trust & 0.856 & 0.889 & 0.906 & 0.710 \\
\rowcolor{gray!25} Motivations & 0.713 & 0.722 & 0.835 & 0.718 \\
Risk tolerance & 0.715 & 0.754 & 0.795 & 0.667 \\
\rowcolor{gray!25} Computer self-efficacy & 0.802 & 0.809 & 0.847 & 0.736 \\
Selective information processing & 0.711 & 0.714 & 0.849 & 0.741 \\
\rowcolor{gray!25} Learning by tinkering & 0.721 & 0.722 & 0.817 & 0.697 \\
Behavioral intention & 0.827 & 0.831 & 0.920 & 0.851 \\ \hline
\end{tabular}
\label{table:internalreliability}
\begin{tablenotes}
\scriptsize
\item Cronbach's $\alpha$ tends to underestimate reliability, whereas composite reliability (CR: $\rho_c$) tends to overestimate it. The true reliability typically lies between these two estimates and is effectively captured by CR($\rho_a$) \cite{russo2021pls}.
\end{tablenotes}
\vspace{-4px}
\end{table}

\textit{1) \textbf{Convergent validity}} examines how a measure correlates with alternate measures of the same construct, focusing on the correlations between indicators (questions) and their corresponding construct. This evaluation assesses whether respondents interpret the questions as intended by the question designers \cite{kock2014advanced}. Our theoretical model comprises latent constructs that are reflectively measured, meaning the changes in the construct should be reflected in changes in the indicators \cite{russo2021pls}. Consequently, these indicators should exhibit a significant proportion of shared variance by converging on their respective constructs \cite{hair2019use}. 
We assessed convergent validity using Average Variance Extracted (AVE) and indicator reliability through outer loadings  \cite{hair2019use}. 
%

AVE represents a construct's communality, indicating the shared variance among its indicators, and should exceed 0.5 \cite{hair2019use}. AVE values for all latent constructs in our model surpassed this threshold (see Table \ref{table:internalreliability}). 
Regarding outer loadings, values above 0.708 are considered sufficient, while values above 0.60 are sufficient for exploratory studies \cite{hair2019use}. We removed variables that did not sufficiently reflect changes in the latent construct (SE3 from computer self-efficacy and IP3 from selective information processing).\footnote{After removing SE3 and IP3, the AVE values for computer self-efficacy (now with 3 indicators) and selective information processing (now with 2 indicators) increased from 0.627 to 0.736 and 0.609 to 0.741, respectively.} Thus, all indicators in our model exceeded the threshold, ranging between 0.615 and 0.954 (see Fig. \ref{fig:model}).

\textit{2) \textbf{Internal consistency reliability}} seeks to confirm that the indicators are consistent with one another and that they consistently and reliably measure the same construct. 
To assess this, we performed both Cronbach's $\alpha$ and Composite Reliability (CR: $\rho_a, \rho_c$) tests \cite{russo2021pls}. The desirable range for these values is between 0.7 and 0.9 \cite{hair2019use}. 
As presented in Table \ref{table:internalreliability}, all values corresponding to our model constructs fall within the acceptable range, confirming that the constructs and their indicators meet the reliability criteria.

\textit{3) \textbf{Discriminant validity}} assesses the distinctiveness of each construct in relation to the others.  
Our model includes 10 latent variables (Table \ref{table:internalreliability}). 
A primary method for assessing discriminant validity is the Heterotrait-Monotrait (HTMT) ratio of correlations \cite{henseler2015new}. Discriminant validity may be considered problematic if the HTMT ratio $>0.9$, with a more conservative cut-off at $0.85$ \cite{hair2019use}. In our case, the HTMT ratios between the latent constructs ranged from 0.064 to 0.791, all below the threshold. We report the HTMT ratios in the supplemental \cite{supplemental}, along with the cross-loadings of the indicators, and the Fornell-Larcker criterion values for the sake of completeness.


\textit{4) \textbf{Collinearity assessment}} is conducted to evaluate the correlation between predictor variables, ensuring they are independent to avoid potential bias in the model path estimations. We assessed collinearity using the Variance Inflation Factor (VIF). In our model, all VIF values are below 2.1, well below the accepted cut-off value of 5 \cite{hair2019use}.

\vspace{-1mm}
\subsection{Structural Model Evaluation}
\label{str-eval}
\vspace{-1mm}

After confirming the constructs' reliability and validity, we assess the structural model (graphically represented in Fig. \ref{fig:model}). This evaluation involves validating the research hypotheses and assessing the model's predictive power.

\subsubsection{\textbf{Path coefficients and significance}}  

Table~\ref{tab:new_path_analysis} presents the results of the hypotheses testing, including the mean of the bootstrap distribution (B), the standard deviation (SD), the 95\% confidence interval (CI), and the p-values. The path coefficients in Fig.~\ref{fig:model} and Table~\ref{tab:new_path_analysis} are interpreted as standard regression coefficients, indicating the direct effects of one variable on another. Each hypothesis is represented by an arrow between constructs in Fig.~\ref{fig:model}. For instance, the arrow from ``Functional Value'' to ``Trust'' corresponds to H2. Given its positive path coefficient (\textit{B}=0.142), genAI's \textit{functional value} is positively associated with developers' \textit{trust} in these tools. The coefficient of 0.142 indicates that when the score for \textit{functional value} increases by one standard deviation unit, the score for \textit{trust} increases by 0.142 standard deviation units. The analysis results (Table \ref{tab:new_path_analysis}) show that most of our hypotheses are supported, except for H3 (p=0.58), H9 (p=0.06), and H10 (p=0.33). Next, we detail the factors associated with trust and behavioral intentions for the supported hypotheses with some exemplary quotes from responses to the open-ended question to illustrate our findings. 

\noindent\textit{\textbf{Factors associated with trust (RQ1)}}: 
Our analysis supported Hypotheses H1 (p=0.00), H2 (p=0.03), and H4 (p=0.00) (Table \ref{tab:new_path_analysis}). 
First, the support for \textit{system/output quality} in fostering trust (H1) can be explained by how developers prefer tools that deliver accurate, reliable outputs matching their work style and practices \cite{wang2023investigating, yu2019trust}. 
Next, the \textit{functional value} of genAI, encompassing educational benefits and practical advantages, promotes trust (H2) since developers prioritize tools that offer tangible utility in their work \cite{johnson2023make, ziegler2024measuring}. For instance, one survey respondent mentioned the practical utility of genAI tools, stating, \textit{``I find value in these models for creative endeavors, gaining different perspectives, or coming up with ideas I wouldn't have otherwise''.}
Finally, \textit{goal maintenance} is relevant for cultivating trust (H4). The alignment between a developer's goals and genAI's actions supports using genAI tools to achieve these goals. This eliminates the need for developers to constantly verify the relevance of genAI's outputs, thereby reducing cognitive load. This congruence ultimately enhances genAI's credibility as a cognitive collaborator \cite{wischnewski2023measuring} rather than as an independent and potentially untrustworthy tool, thus bolstering trust in these tools. 

\noindent\textit{\textbf{Factors associated with behavioral intentions (RQ2)}}: Our analysis supported Hypotheses H5 (p=0.00), H6 (p=0.01), H7 (p=0.01), and H8 (p=0.00), indicating that developers' trust (H5) and their cognitive styles—motivations (H6), computer self-efficacy (H7), and risk tolerance (H8)—are significantly associated with their behavioral intentions to use genAI tools. 

Trust (H5) is pivotal in shaping adoption decisions as it reduces resistance to new technologies \cite{xiao2014social, witschey2015quantifying}. When developers trust genAI tools, they perceive them as credible partners, enhancing their willingness to use these tools. Moreover, developers' cognitive styles significantly shape their intentions to adopt genAI tools. Developers motivated by the intrinsic enjoyment of technology (H6) have higher intentions to adopt genAI tools. In contrast, those with a task-oriented approach tend to be more cautious and hesitant about the cognitive effort they are willing to invest in these tools \cite{burnett2016gendermag}. One respondent echoed this, stating, \textit{``I am slow to adopt new workflows; I put off actively exploring new tools unless it is related to what I need to do''.} Higher computer self-efficacy within peer groups is also significantly associated with increased intentions to use genAI tools (H7). Despite generally high self-efficacy, some developers face interaction challenges with genAI that may impact their confidence and adoption rates. One respondent shared, \textit{``I see my colleagues getting good responses while I fiddle around to get the answers I need. Also, it does not always show the right document relevant to me, so I prefer traditional ways''.} Furthermore,
we found that developers with higher risk tolerance are significantly more inclined to use these tools than risk-averse individuals (H8). The context (and involved stakes) in which these tools are used further play a role, as highlighted by another respondent: \textit{``I don't use it yet to write code that I can put my name behind in production; I just use it for side projects or little scripts to speed up my job, but not in actual production code''.} 

Finally, our analysis supported Hypothesis H11 (p=0.00), highlighting a significant positive association between developers' behavioral intention to use genAI tools and its usage in their work. This corroborates with prior technology acceptance models \cite{venkatesh2003user, venkatesh2012consumer}, emphasizing the pivotal role of behavioral intentions in predicting use behavior.

\begin{table}[!t]
\centering
\caption{Standarized path coefficients (B), standard deviations (SD), confidence intervals (CI), p values, and effect sizes ($f^2$)}
\label{tab:new_path_analysis}
\vspace{-8px}
\robustify{\bfseries}
\sisetup{
    mode=text,
    group-digits = false ,
    input-signs ={-},
    input-symbols = ( ) [ ] - + *,
    detect-weight=true, 
    detect-family=true,
    table-format=0.2,
    add-decimal-zero=false, 
    add-integer-zero=false,
    round-mode=places, 
    round-precision=2, 
    parse-numbers = true
}
\begin{tabular}{P{2.9cm}  
                S
                S[table-format=0.2]
                >{\centering\arraybackslash}p{1.3cm}
                S[table-format=0.3,round-precision=3]
                S[table-format=0.3,round-precision=2]} 
\toprule
& {\textit{B}} & {SD} & {95\% CI} & {\textit{p}} & {$f^2$} \\
\midrule \midrule
\rowcolor{gray!25} \hangindent1em H1 System/Output quality$\rightarrow$Trust & .596 & .602 & (.45, .77) & \bfseries .000 & .458\\
\hangindent1em H2 Functional value$\rightarrow$Trust & .142 & .065 & (.01, .26) & \bfseries .029 & .171\\
\rowcolor{gray!25} \hangindent1em H3 Ease of use$\rightarrow$Trust & .023 & .064 & (-.08, .16) & .588 & 0.01\\
\hangindent1em H4 Goal maintenance $\rightarrow$Trust & .237 & .074 & (.07, .36) & \bfseries .002 & 0.186\\
\midrule
\rowcolor{gray!25} \hangindent1em H5 Trust$\rightarrow$BI & .485 & .052 & (.38, .58) &  \bfseries .000 & 0.539\\
\hangindent1em H6 Motivations$\rightarrow$BI & .231 & .08 & (.07, .40) &  \bfseries .005 & 0.263\\
\rowcolor{gray!25} \hangindent1em H7 Computer self-efficacy$\rightarrow$BI & .135 & .052 & (.03, .24) &  \bfseries .012 & 0.141\\
\hangindent1em H8 Risk tolerance$\rightarrow$BI & .204 & .061 & (.09, .33) &  \bfseries .001 & 0.181\\
\rowcolor{gray!25} \hangindent1em H9 Selective information processing $\rightarrow$BI & .090 & .046 & (-.02, .14) &  .065 & 0.025\\
\hangindent1em H10 Learning by tinkering$\rightarrow$BI & 0.062 & .061 & (-.06, .18) &  .331 & 0.035\\ 
\rowcolor{gray!25} \hangindent1em H11 BI$\rightarrow$Usage & .567 & .051 & (.46, .67) &  \bfseries .000 & 0.447\\ 
\midrule
\hangindent1em SE Experience$\rightarrow$Trust & .075 & .049 & (-.01, 0.2) &  .125 & 0.016\\  
\rowcolor{gray!25} \hangindent1em SE Experience$\rightarrow$BI & 0.018 & .04 & (-.03, .11) &  .332 & 0.004\\ 
\hangindent1em SE Experience$\rightarrow$Usage & 0.054 & .051 & (-.06, .15) &  .275 & 0.005\\ 

\bottomrule

\end{tabular}
\begin{tablenotes}
  \scriptsize
\item BI: Behavioral Intention. We consider $f^2<$ 0.02 to be no effect, $f^2 \in$ [0.02, 0.15) to be small, $f^2 \in$ [0.15, 0.35) to be medium, and $f^2 >$ 0.35 to be large \cite{cohen2013statistical}.
\end{tablenotes}
\vspace{-35px}
\end{table}

\noindent\textit{\textbf{Control variables}}: Although experience is often relevant for technology adoption \cite{venkatesh2003user}, our analysis found no significant associations between SE experience and trust, behavioral intentions, or usage of genAI tools. This is likely since genAI introduces a new interaction paradigm \cite{weisz2023toward}, which diverges from traditional SE tools and requires different skills and interactions not necessarily linked to SE experience.  
%
Familiarity with genAI, while potentially influential, was excluded as a control variable due to a highly skewed distribution of responses, with most participants reporting high familiarity. Including such skewed variables could lead to unreliable estimates and compromise the model's validity \cite{hair2019use, sarstedt2019partial}. Similarly, the gender variable was excluded due to its skewed distribution.  
The analysis of \textit{unobserved heterogeneity} (see supplemental \cite{supplemental}) confirms the absence of any group differences in the model (e.g., organizational heterogeneity) caused by unmeasured criteria.

\vspace{-2mm}
\subsubsection{\textbf{Model evaluation}} We assessed the relationship between constructs and the predictive capabilities of the theoretical model by evaluating the model’s explanatory power ($R^2$, Adjusted (Adj.) $R^2$), model fit (SRMR), effect sizes ($f^2$), and predictive relevance ($Q^2$) \cite{russo2021pls}.

\textit{Explanatory power}: The coefficient of determination ($R^2$ and Adj. $R^2$ values)
indicate the proportion of variance in the endogenous
variables explained by the predictors. Ranging from 0 to 1, higher $R^2$ values signify greater explanatory power, with 0.25, 0.5, and 0.75 representing weak, moderate, and substantial levels, respectively \cite{hair2019use}. As shown in Table \ref{tab:r2-table}, the $R^2$ values in our model are 0.68 for Trust, 0.66 for Behavioral intention, and 0.33 for Usage, demonstrating moderate to substantial explanatory power, well above the accepted threshold of 0.19 \cite{chin1998partial}. Further, Table \ref{tab:new_path_analysis} presents the effect sizes ($f^2$), which measure the impact of each predictor on the endogenous variables. The effect sizes indicate that the predictors exhibit medium to large effects on their respective endogenous variables for all supported hypotheses in our model, with values ranging from 0.14 to 0.54 \cite{cohen2013statistical}, further corroborating the model's explanatory power.\footnote{Large $R^2$ and $f^2$ can occasionally indicate overfitting. We thoroughly evaluated this issue by analyzing residuals and conducting cross-validation, finding no evidence of model overfitting (see  \cite{supplemental}).}

\textit{Model fit:} We analyzed the overall model fit using the standardized root mean square residual (SRMR), a recommended fit measure for detecting misspecification in PLS-SEM models \cite{russo2021pls}. Our results suggest a good fit of the data in the theoretical model, with SRMR = 0.077, which is below the suggested thresholds of 0.08 (conservative) and 0.10 (lenient) \cite{henseler2016using}.

\textit{Predictive relevance:} Finally, we evaluated the model’s predictive relevance using Stone-Geisser’s $Q^2$ \cite{stone1974cross}, a measure of external validity \cite{hair2019use} obtainable via the PLS-predict algorithm \cite{shmueli2016elephant} in SmartPLS. PLS-predict is a holdout sample-based procedure: it divides the data into $k$ subgroups (folds) of roughly equal size, using ($k-1$) folds as a training sample to estimate the model, while the remaining fold serves as a holdout to assess out-of-sample predictive power.
$Q^2_\text{predict}$ values are calculated for endogenous variables; values greater than 0 indicate predictive relevance, while negative values suggest the model does not outperform a simple average of the endogenous variable. Our sample was segmented into k=10 parts, and 10 repetitions were used to derive the $Q^2_\text{predict}$ statistic \cite{hair2019use}, all of which were greater than 0 (Table \ref{tab:r2-table}), confirming our model's adequacy in terms of predictive relevance.

\vspace{-3mm}
\begin{table}[htb]
\caption{Coefficient of determination and predictive relevance}
\label{tab:r2-table}
\vspace{-8px}
\centering
\begin{tabular}{P{3cm}  
                S[table-format=1.4]  
                S[table-format=1.4]  
                S[table-format=1.4]} 
\toprule
\textbf{Construct} & \textit{$R^2$} & \textit{Adj. $R^2$} & \textit{$Q^2_\text{predict}$} \\
\midrule
\rowcolor{gray!25} Trust & 0.679 & 0.672 & 0.679 \\
Behavioral Intention & 0.658 & 0.647  & 0.648 \\
\rowcolor{gray!25} Usage & 0.331 & 0.326 & 0.234 \\
\bottomrule
\end{tabular}
\vspace{-2.5mm}
\end{table}

\subsubsection{\textbf{Common method bias}} We collected data via a single survey instrument, which might raise concerns about Common Method Bias/Variance (CMB/CMV) \cite{russo2021pls}. To test for CMB, we applied Harman’s single factor test \cite{podsakoff2003common} on the latent variables. No single factor explained more than 23\% variance. An unrotated exploratory factor analysis with a forced single-factor solution was conducted, which explained 30.3\% of the variance, well below the 50\% threshold. Additionally, we used Kock’s collinearity approach \cite{kock2015common}. The VIFs for the latent variables ranged from 1.01 to 2.45, all under the cut-off of 3.3. These indicate that CMB was not a concern in our study.

\vspace{-2mm}
\vspace{1.5mm}
\section{Discussion}
\label{sec:discussion}



\subsection{Implications for practice}
\label{impl-prac}

{\textbf{\textit{Design to maintain developers' goals.}}}
Our findings suggest that developers' trust in genAI tools manifests when these tools align with their goals (H4). This is likely because goal maintenance reduces the need for developers to constantly verify the relevance of genAI's contributions, easing cognitive load and allowing them to focus on task-related activities. 

To achieve goal maintenance, the AI should consistently account for the developer's \textbf{(1)} \textit{current state}; \textbf{(2)} \textit{immediate goals}, and the \textit{expected success} (outcomes from AI) \cite{wischnewski2023measuring}; as well as \textbf{(3)} \textit{preferences} for transitioning from their current state (1) to achieving their immediate goals (2). These preferences may involve process methodologies, interaction styles, output specifications, alignment with the work style, or safety/security considerations \cite{johnson2023make, burnett2016gendermag}. 
Given that an individual's goal(s) may evolve in response to changes in their task state, toolsmiths must also prioritize \textit{adaptability} in genAI tool design. For instance, adaptability can be incorporated by iteratively adjusting AI's actions based on user input on (1), (2), and (3) to ensure its ongoing alignment with their shifting goals.   


Allowing developers the flexibility to \textit{explicitly steer AI's actions} as needed is also important for goal maintenance. This control can be essential if the genAI tool deviates from the expected trajectory, enabling developers to (re)calibrate it to support their goals. 
Further, this can also support developers' metacognitive flexibility \cite{tankelevitch2024metacognitive}, i.e., they may adapt their cognitive strategies based on new information or task-state changes. The ability to steer the AI to align with these adapted strategies helps accommodate their evolving goals.


\textbf{\textit{Design for contextual transparency.}}
Developers often face decisions about incorporating genAI tool support for their tasks. 
Any mismatch between their expectations and genAI's true capabilities can lead to over/under-estimating the tool's functionality, increasing the risk of errors, lost productivity, or potential adverse outcomes in critical tasks \cite{pearce2022asleep}. Such outcomes are detrimental to fostering subsequent trust. Thus, helping developers to calibrate their expectations to match the true tool capabilities is essential. 
Creating this alignment involves designing interfaces that \textit{explicitly communicate the tool's capabilities and limitations}, consistent with HAI interaction guidelines \cite{amershi2019guidelines, GoogleGuidelines}. This clarity allows individuals to form accurate expectations about system quality (H1) within their task contexts and assess its functional value (H2), thereby fostering appropriate trust \cite{jacovi2021formalizing}. 
To that end, we suggest:


\noindent(a) \textit{Communicating genAI's reasoning process and derivation of its outputs in the context of the current task}. Such transparency about the source and how an AI arrived at its output can enable developers to assess correctness more accurately and evaluate any safety or security issues related to using these outputs in their work. This can enhance the assessment of system/output quality, thus cultivating appropriate trust (H1).

\noindent(b) \textit{Communicating genAI's limitations for the current task and scoping assistance under conditions of uncertainty to incite warranted trust} \cite{jacovi2021formalizing}.
Doing so allows developers to consider the tool’s practical utility for their task context, clarifying its appropriate functional value (H2) in their work.


\textbf{\textit{Inclusive tool design for HAI-UX fairness.}} 
AI fairness has gained substantial traction over the years \cite{green2019disparate}. While much of the research and discussion has focused on data or algorithmic fairness, fairness should also include user experiences in human-AI interactions (HAI-UX). We advocate promoting fairness in HAI-UX through inclusive genAI tool design, specifically by supporting developers' diverse cognitive styles.

Our findings indicate that developers who are motivated to use genAI tools for their own sake (H6), have higher computer self-efficacy (H7), and have greater tolerance for risk (H8) are more inclined to adopt these tools compared to their peers. To achieve more equitable acceptance of these tools across the cognitive spectrum of developers, future designs must prioritize \textit{adaptability} based not only on developers' goals (discussed above) but also on their cognitive styles. This can be achieved by capturing these styles (e.g., using the survey questions from this study) and designing genAI tools that dynamically adapt to align with these styles using various strategies. For example, to support developers motivated by task completion, genAI tools can solicit their immediate goals and expected outcomes and deliver contextually appropriate information (and explanations) consistent with their preferred styles. This would simplify developers' ongoing tasks without overwhelming them with unrelated features or extraneous information, helping them complete their tasks effectively.
As another example, genAI tools can support individuals with lower self-efficacy by offering explicit cues that differentiate between errors arising from prompting issues and those due to system limitations. For instance, it can caution users about tasks where it typically underperforms, based on prior feedback, or highlight specific parts of the prompt that influenced the generated output. This distinction can help prevent individuals
from doubting their ability to effectively use these tools in their work.
Further, for risk-averse developers, transparency in genAI behavior and outputs should be emphasized. GenAI should also provide explicit indicators of any potential flaws or uncertainties in its outputs (e.g., confidence scores \cite{wang2023investigating}). Such transparency can help developers make informed decisions about incorporating these outputs into their work (e.g., using AI-generated code), accommodating the caution and deliberation associated with risk aversion.
\vspace{-1.5mm}
\subsection{Implications for research}
\label{impl-res}
Our study establishes an understanding of developers' trust and intention-related factors during the early stage of genAI adoption. 
Furthermore, our study offers a validated instrument for capturing relevant factors in the context of human-genAI interaction. Researchers can utilize this instrument to operationalize theoretical expectations or hypotheses—such as capturing the dynamics of trust and intentions in finer contexts, refining genAI tools with design improvements, and comparing user experiences before and after design changes; thus advancing the understanding of AI adoption.

\textit{\textbf{Non-significant associations}}: Our analysis did not find support for Hypotheses H3 (p=0.59), H9 (p=0.06), and H10 (p=0.33). 
These findings are surprising, as ease of use, information processing, and tinkering learning style are relevant when considering traditional software tools \cite{burnett2016gendermag, venkatesh2003user}. However, in genAI contexts, these constructs may manifest differently due to the altered dynamics of user engagement compared to more traditional software. The intuitive nature of genAI interfaces might diminish the traditional impact of these factors. 
For example, ease of use might not show a relation as using these interfaces is inherently easy; instead, the appropriateness of the queries is what matters.
Similarly, developers' information processing style (H9) did not significantly influence their intentions to use genAI tools, likely because how individuals articulate their needs---a single comprehensive prompt or sequence of queries---often aligns with their preferences for consuming information (comprehensive or selective). 
The lack of a relationship for tinkering style (H10), as well, could be attributed to genAI's interaction paradigm, which is primarily centered around (re)formulating and following up with queries rather than ``tinkering'' with the software's features.
If these speculations hold, how certain validated constructs were framed in the current study \cite{supplemental} might have indeed limited our understanding of these dynamics. 
Future research should explore these constructs more deeply within the context of human-genAI interactions. For instance, instead of focusing on `ease of use' or `tinkering with software features', studies could examine `ease of prompting' or `tinkering with prompt strategies' and how preferences (and proficiency) in these areas influence developers' trust and behavioral intentions. Understanding these dynamics can inform the future design and adoption strategies of genAI tools, aligning them more closely with user interaction patterns and cognitive styles.

\subsection{Threats to validity and limitations}
\label{sec:threat}

\textbf{\textit{Construct validity}}: We captured constructs through self-reported measures, asking participants to express their agreement with literature-derived indicators. This approach assumes that participants' responses accurately reflect their beliefs and experiences, which might not always be the case due to various biases or misinterpretations. To reduce this threat, we used validated instruments, evaluated the psychometric validity of the PICSE questions, involved practitioners in designing the questions, ran pilot studies, incorporated attention checks, randomized the questionnaire within blocks, and screened the responses. Further, our analysis confirmed that the constructs were internally consistent, reliable, and met convergent and discriminant validity criteria. 
We did not directly ask participants about their genAI experience; instead, we used their familiarity with genAI tools and its frequency of use at work as proxies to capture their experience with these tools.


\textbf{\textit{Internal validity}}: Our hypotheses propose associations between constructs, rather than causal relationships, given the cross-sectional nature of the study \cite{stol2018abc}. We acknowledge the limitation of self-selection bias, as respondents interested in (or skeptical about) genAI tools might be more willing to complete the questionnaire.
Further, a theoretical model like ours cannot capture an exhaustive list of factors. Other factors can play a role, thus positioning our results as a reference for future studies. Future work should also consider using longitudinal data and control for potential confounding factors, such as familiarity with genAI and demographic variables. 
Additionally, trust is a situation-dependent construct \cite{jacovi2021formalizing}. Although we focused on software development, trust in genAI tools may vary based on finer work contexts (e.g., software design vs. software testing tasks). Therefore, our results should be interpreted as a theoretical starting point, guiding future studies to explore these contextual influences.

\textbf{\textit{External validity}}:
Our survey was conducted within GitHub and Microsoft. While the sample includes engineers from around the globe, it may not fully represent the broader software developer community. However, the sample distribution aligns with previous empirical studies involving software engineers \cite{russo2024navigating, trinkenreich2023belong}, providing a suitable starting point to understand the associations presented in our model. 
The responses were sufficiently consistent to find full or partial empirical support for the hypotheses. 
Nevertheless, theory development is an iterative process \cite{shull2007guide}. Thus, our results should be interpreted as a starting point, aiming for theoretical rather than statistical generalizability. Future studies should replicate, validate, and extend our theoretical model in various contexts.
\vspace{-2mm}
\vspace{0.5mm}
\section{Conclusion}
\label{sec:conclusion}

Our findings highlight that genAI's system/output quality, functional value, and goal maintenance significantly influence developers' trust. Furthermore, trust and cognitive styles (motivations, computer self-efficacy, and attitude towards risk) influence intentions to use these tools, which, in turn, drives usage. We also contribute a validated instrument to capture trust-related factors in human-genAI interaction contexts.

Beyond theoretical contributions, our study offers practical implications, including pointers for design mechanisms that foster appropriate trust and promote equitable user experiences (Sec. \ref{sec:discussion}).
While our work enhances the understanding of developers' trust and behavioral intentions towards genAI tools, long-term longitudinal studies are essential for refining the knowledge of AI adoption dynamics in software engineering.


\section*{Acknowledgments} 

We thank the GitHub Next team, Tom Zimmermann, and Christian Bird for providing valuable feedback on the survey contents and Brian Houck for facilitating the survey distribution. We also thank all the survey respondents for their time and insights. This work was partially supported by the National Science Foundation under Grant Numbers 2235601, 2236198, 2247929, 2303042, and 2303043. Any opinions, findings, conclusions, or recommendations expressed in this material are those of the authors and do not necessarily reflect the views of the sponsors.


\bibliographystyle{IEEEtranN}
\footnotesize{\bibliography{acmart}}

\end{document}